\documentclass[runningheads]{llncs}
\usepackage[T1]{fontenc}
\usepackage{enumitem}
\usepackage{graphicx}
\usepackage{color, colortbl}
\definecolor{Gray}{gray}{0.9}
\usepackage{graphicx}
\usepackage{xcolor}
\usepackage{makecell}
\usepackage{url}
\usepackage{adjustbox}

\usepackage{amsmath}
\usepackage{amssymb}
\usepackage{lipsum}
\usepackage{subfig}
\usepackage{booktabs}
\usepackage{pifont}
\usepackage{tabularx,booktabs}
\usepackage{multirow}
\usepackage{mathrsfs}

\newcommand{\cmark}{\ding{51}}%
\newcommand{\xmark}{\ding{55}}%
\usepackage[acronym,nomain]{glossaries}

\newacronym{mr}{MR}{Magnetic Resonance}
\newacronym{nor}{NOR}{normal subjects}
\newacronym{minf}{MINF}{previous myocardial infarction}
\newacronym{dcm}{DCM}{dilated cardiomyopathy}
\newacronym{hcm}{HCM}{hypertrophic cardiomyopathy}
\newacronym{arv}{aRV}{abnormal right ventricle}

\newacronym{rv}{RV}{right ventricle}
\newacronym{lv}{LV}{left ventricle}
\newacronym{vae}{VAE}{Variational AutoEncoder}
\newacronym{sivae}{SIVAE}{Soft Introspective Variational AutoEncoder}
\newacronym{arsivae}{AR-SIVAE}{Attribute Regularized Soft Introspective Variational Autoencoder}
\newacronym{gan}{GANs}{Generative Adversarial Networks}
\newacronym{mse}{MSE}{Mean Squared Error}
\newacronym{lvedv}{LVEDV}{Left Ventricular End-Diastolic Volume}
\newacronym{myoedv}{MEDV}{Myocardial End-Diastolic Volume}
\newacronym{rvedv}{RVEDV}{Right Ventricular End-Diastolic Volume}
\newacronym{ssim}{SSIM}{Structural Similarity Index Measure}
\newacronym{psnr}{PSNR}{Peak Signal-to-Noise Ratio}
\newacronym{sap}{SAP}{Separated Attribute Predictability}
\newacronym{scc}{SCC}{Spearman Correlation Coefficient}
\newacronym{mlp}{MLP}{Multi Layer Perceptron}
\newacronym{xai}{XAI}{explainable AI}
\newacronym{elbo}{ELBO}{Evidence Lower Bound}
\newacronym{lpips}{LPIPS}{Learned Perceptual Image Patch Similarity}
\newacronym{ed}{ED}{end-diastole}
\newacronym{es}{ES}{end-systole}

\begin{document}
\title{Interpretable Representation Learning of Cardiac MRI via Attribute Regularization}
%
%
\author{Maxime Di Folco\inst{1,2} \and
Cosmin I. Bercea \inst{1,2,3} \and
Emily Chan \inst{1,2} \and
Julia A. Schnabel \inst{1,2,3,4}}

%
\authorrunning{M. Di Folco et al.}


\institute{Institute of Machine Learning in Biomedical Imaging, Helmholtz  Munich \and
 Helmholtz AI, Helmholtz  Munich \\
 \and School of Computation, Information and Technology, Technical University of Munich \and
School of Biomedical Engineering and Imaging Sciences, King's College London\\
\email{maxime.difolco@helmholtz-munich.de}}

\maketitle 
\begin{abstract}

Interpretability is essential in medical imaging to ensure that clinicians can comprehend and trust artificial intelligence models. Several approaches have been recently considered to encode attributes in the latent space to enhance its interpretability. Notably, attribute regularization aims to encode a set of attributes along the dimensions of a latent representation. However, this approach is based on Variational AutoEncoder and suffers from blurry reconstruction. In this paper, we propose an Attributed-regularized Soft Introspective Variational Autoencoder that combines attribute regularization of the latent space within the framework of an adversarially trained variational autoencoder. We demonstrate on short-axis cardiac Magnetic Resonance images of the UK Biobank the ability of the proposed method to address blurry reconstruction issues of variational autoencoder methods while preserving the latent space interpretability.

\end{abstract}
\section{Introduction} 

\begin{figure}[t!]
    \centering
    \includegraphics[width=1\textwidth]{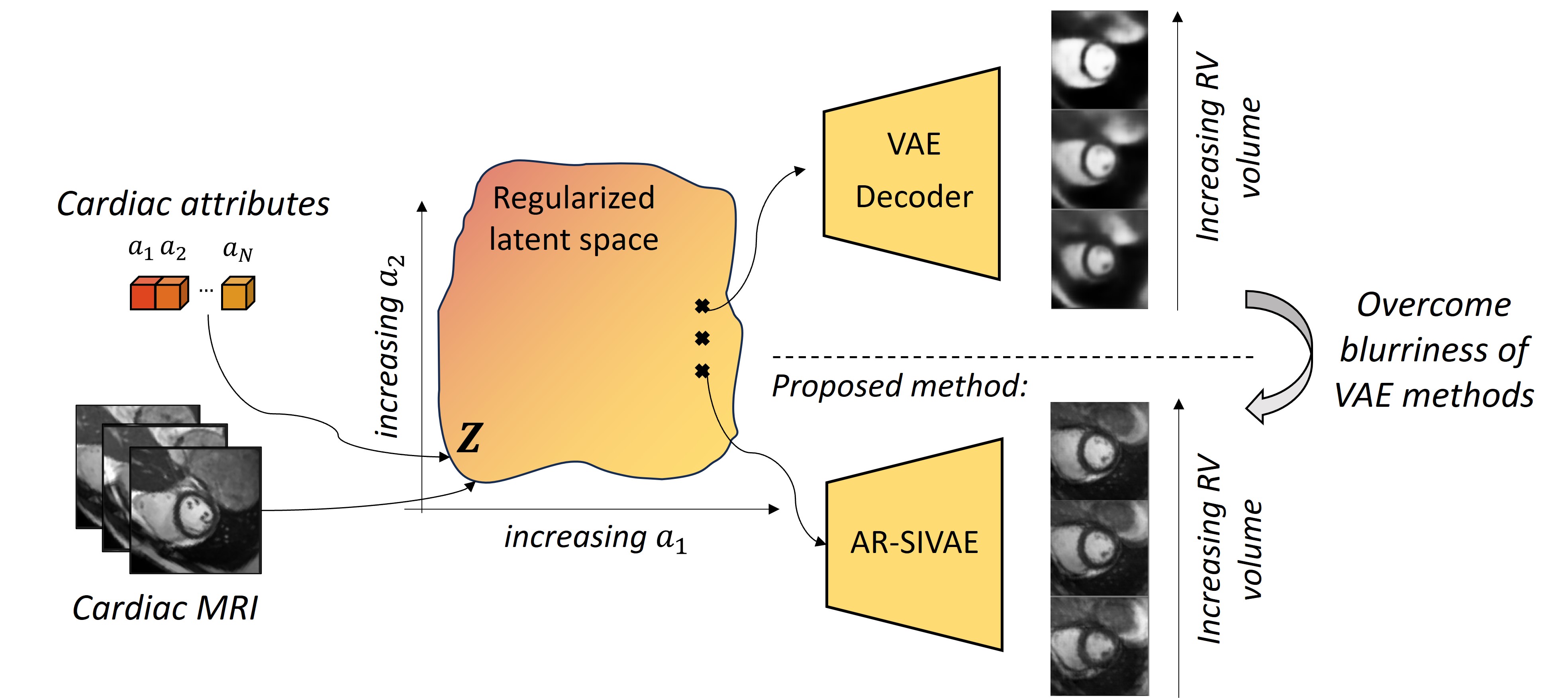}
    \caption{\textbf{AR-SIVAE}: Attribute regularized Soft Introspective Variational AutoEncoder (AR-SIVAE) combines attribute regularization within the SIVAE framework (an adversarially trained encoder-decoder network) to enhance the interpretability of the latent space while being able to generate non-blurry samples.}
    \label{fig:enter-label}
\end{figure}

Interpretability is crucial for transparent AI systems in medical imaging to build clinician trust and advance AI adoption in clinical workflows. As highlighted by Rudin \cite{Rudin:2019}, it is essential that models are inherently interpretable and not black-box models to ensure their relevance.
Latent representation models like \gls{vae} have emerged as potent models capable of encoding crucial hidden variables within the input data \cite{higgins:2017,biffi:2020,Liu:2020}. Especially when dealing with data that contain different interpretable features (data attributes), self-supervised \cite{Hager:2023} or supervised approaches can encode those attributes in the latent space \cite{Mirza:2014,Fader:2017,engel2017latent,Pati:2021}.

In this context, Pati et al. \cite{Pati:2021} introduced an attribute-regularized method based on VAEs that aims to regularize each attribute, added as extra input, along a dimension of the latent space and, therefore, ensure the latent space interpretability. Notably, Cetin et al. \cite{Cetin:2023} applied this architecture for cardiac attributes on MRI data, demonstrating a significant improvement in the interpretability of the latent representation and its relevance for a downstream cardiac disease classification task. Nevertheless, VAE-based methods may suffer from blurry reconstruction, which could be problematic for any downstream task. In order to overcome this limitation while preserving the latent interpretability of VAEs, Daniel et al.~\cite{Daniel:2021} introduced a novel approach called \gls{sivae}. SIVAEs leverage the benefits of  VAEs and \gls{gan} by incorporating an adversarial loss into VAE training. In contrast to earlier methods that used additional discriminator networks \cite{pidhorskyi2018generative}, SIVAE utilizes the encoder and decoder of VAE in an adversarial manner.

In this paper, we propose the Attributed-regularized Soft Introspective Variational Autoencoder (AR-SIVAE) by combining an attribute regularization loss in the \gls{sivae} framework to preserve the interpretability of the latent space while having better image generation capabilities. To the best of our knowledge, we are the first to introduce this loss in an adversarially trained \gls{vae}. We compare our method to the one described in Cetin et al. \cite{Cetin:2023} on a healthy population of cardiac MRI from the UK Biobank. Our method overcomes the limitations associated with blurry reconstruction while maintaining latent space interpretability.

\section{Methods}

\subsection{Preliminaries:}
\label{sec:rec}

\subsubsection{Attributed regularized VAE}: Attri-VAE, proposed by Pati et al. \cite{Pati:2021}, aims to encode an attribute $a$ along a dimension $k$ of a $\mathbb{D}$-dimensional latent space \begin{math}\mathbf{z}: {z^k}, k \in [0,\mathbb{D}) \end{math}, such that the attribute value increases when we traverse the dimension $k$. This is achieved by adding to the VAE training objective. This attribute regularization loss based on an attribute distance matrix $D_a$ and a similar distance matrix $D_k$ computed from the regularized dimension $k$. They are defined as follows and computed for each batch of the training data:

\begin{equation}
	\centering
	D_a(i,j) = a(x_i) - a(x_j) \text{;  } D_k(i,j) = z_i^k - z_j^k
\end{equation}

where $\mathbf{x_i}, \mathbf{x_j} \in \mathbb{R^N}$ are two high-dimensional samples of dimension $N$ (with $N >> \mathbb{D}$). 
The attribute regularization loss term is then computed for each attribute $k$ as follows:

\begin{equation}
	\centering
	L_{k,a} = MAE(tanh(\delta D_k) - sgn(D_a))
	\label{eq:attr}
\end{equation}

where $MAE(.)$ is the mean absolute error, $tanh(.)$ is the hyperbole tangent function, $sgn(.)$ is the sign function and $\delta$ is a tunable hyperparameter which decide the spread of the posterior distribution. The sum for each attribute is then added to the $\beta$-VAE loss term and weighted by the hyperparameter $\gamma_{reg}$:

\begin{flalign}
     \mathcal{L} =  \mathcal{L}_r(x) + \beta \mathcal{L}_{KL}(x) + \gamma_{reg} \mathcal{L}_{attr} \hspace{2mm} \text{with }  \mathcal{L}_{attr} = \sum^{\mathbb{D}-1}_{l=0}{L_{k_l,a_l}} 
\end{flalign}

where ${\mathscr{A}}: a_l$, $l \in [0,D)$ a set of attributes, $\mathcal{L}_r$ the reconstruction loss and $\mathcal{L}_{KL}$ the Kullback Leibler divergence controlled by the parameter $\beta$.

\subsubsection{Soft Introspective Variational Autoencoder (SIVAE)}

The \gls*{sivae} framework proposed by \cite{Daniel:2021} is an adversarially trained \gls{vae}. Its encoder is trained to distinguish between real and generated samples by minimizing the KL divergence between the latent distribution of real samples and the prior while maximizing the KL divergence of generated samples. Conversely, the decoder aims to deceive the encoder by reconstructing real data samples using the standard \gls{elbo} and minimizing the KL divergence of generated samples embedded by the encoder. The optimization objectives for the encoder, $E_{\Phi}$, and decoder, $D_{\theta}$, to be maximized are formulated as follows:
\begin{flalign}
	\label{eq::sivae}
	& \mathcal{L}_{E_{\phi}}(x,z) = ELBO(x) - \frac{1}{\alpha}(exp(\alpha ELBO(D_\theta(\mathbf{z})),\\
	& \mathcal{L}_{D_{\theta}}(x,z) = ELBO(x) + \eta ELBO(D_\theta(z)) \nonumber
\end{flalign}
where $\alpha \geq 0$ and $\eta \geq 0$ are hyper-parameters.


\begin{figure}[t!]
    \centering
    \includegraphics[width = 1\linewidth]{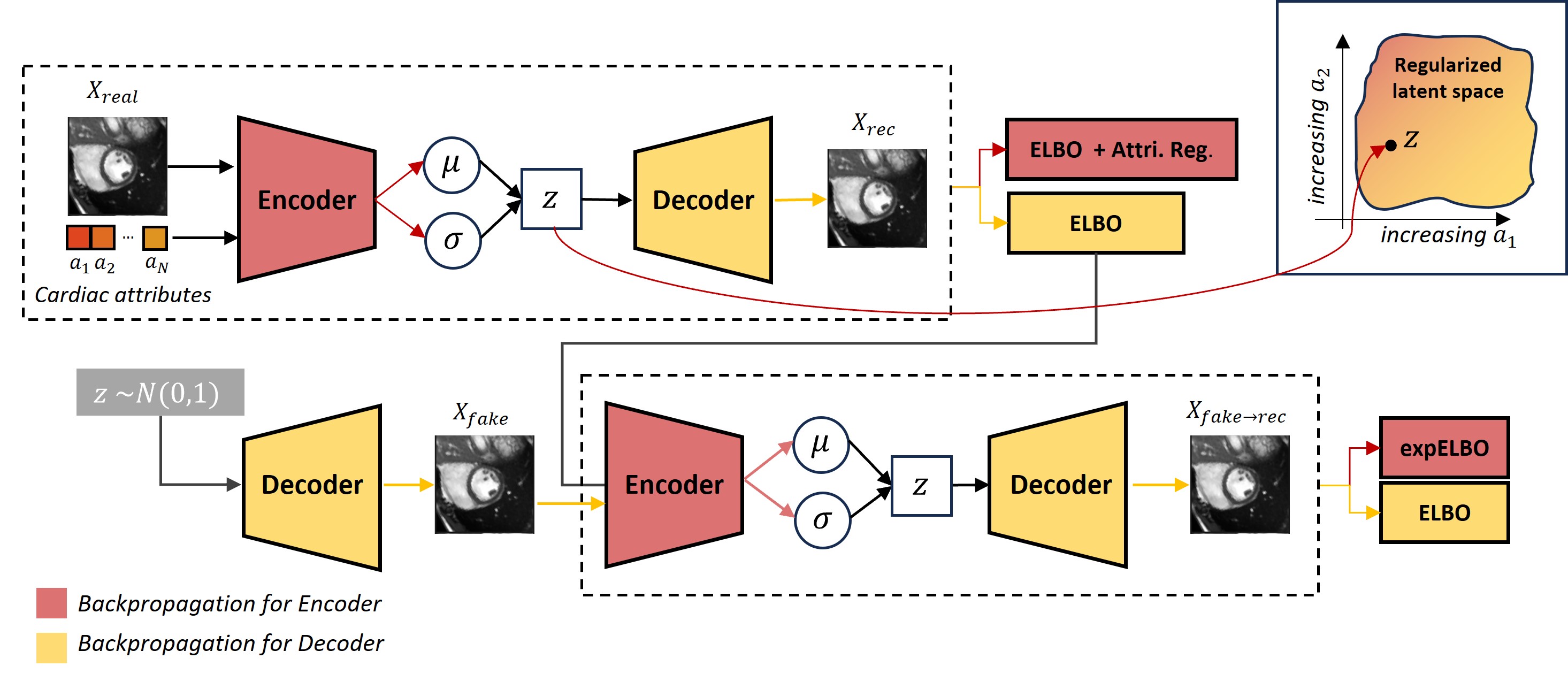}
    \caption{Illustration of the AR-SIVAE framework. Attri. Reg: Attribute regularization.} 
    \label{fig:framework}
\end{figure}

\subsection{Proposed method: AR-SIVAE}
\label{sec:ours}

In this work, we propose an \gls{arsivae} (Fig. \ref{fig:framework}) by adding the attribute regularization loss defined in Eq. \ref{eq:attr} to the encoder loss of the \gls{vae}. The optimization objective of the encoder becomes: 
\begin{flalign}
	\label{eq:OURS}
	\mathcal{L}_{E_{\phi}}(x,z) = \gamma_{reg} \mathcal{L}_{attr} +  ELBO(x) - \frac{1}{\alpha}(exp(\alpha ELBO(D_\theta(z)) 
\end{flalign}

where $\gamma_{reg}$ is a hyperparameter that weights the attribute regularization loss term. Accordingly to the guidelines described in \cite{Daniel:2021}, we always set $\alpha=2$. 
In pratice, the optimization objectives are computed as follows: 

\begin{flalign}
    \mathcal{L}_{E_{\phi}}(x,z) & =  \gamma_{reg} \mathcal{L}_{attr} + s \cdot (\beta_{rec} \mathcal{L}_r(x) + \beta_{kl}\mathcal{L}_{KL}(x)) \\
    & +  \frac{1}{2} \exp(-2s \cdot (\beta_{rec} \mathcal{L}_r(D_\theta(z)) + \beta_{neg}\mathcal{L}_{KL}(D_\theta(z)))) \nonumber \\ 
    \mathcal{L}_{D_{\theta}}(x,z)&  =   s \cdot \beta_{rec} \mathcal{L}_r(x) + s \cdot  (\eta \cdot \beta_{rec} \mathcal{L}_r(D_\theta(z)) + \beta_{kl}\mathcal{L}_{KL}(D_\theta(z))) 
\end{flalign}

where $s$ is a normalizing constant set to the size of the images. $\mathcal{L}_r$ is the reconstruction loss computed as a combination of the \gls{mse} and perceptual loss weighted by an hyperparemeter $\alpha_{pl}$. The training process, detailed in \cite{Daniel:2021}, is composed of two steps: first, the decoder is frozen, and the encoder is updated, and then the encoder is frozen, and the decoder is updated. 

\subsection{Dataset and Implementation details}

\subsubsection{Dataset:} In this work, we processed the Cine MRI acquisitions of the short-axis view of the UK Biobank study \cite{petersen_CMR:2016}. We selected all of the 5392 subjects who self-declared to not have any cardiovascular disease using the UK Biobank field 20002. We preserve only the 5360 cases where the segmentation mask contains at \gls{ed} and \gls{es} the following: more than 10 pixels per region (\gls{lv}, \gls{rv} and myocardium), more than six slices segmented with no discontinuity (no missing segmentation between the slices) and the mid-cavity slice has the \gls{lv} and the \gls{rv} segmented (similar to \cite{bai2018:automated} and detailed in the associated code). Per subject, we selected the basal slice at \gls{ed} and \gls{es}. The barycenter of the left ventricle was centred for each image and was aligned with the right ventricular barycenter along the horizontal axis. The images were cropped around the centre at a size of 128x128 pixels. The dataset was split into 3752, 804 and 804 subjects for training, validation and testing, respectively. 

\subsubsection{Attributes: }
We computed cardiac morphometric attributes for the regularization based on the methods described in \cite{bai2018:automated} and using the associated public code \footnote{\url{https://github.com/baiwenjia/ukbb_cardiac/}}. We used the volume at \gls{ed} and \gls{es} of the \gls{lv}, \gls{rv} and the myocardium denoted LVEDV, RVEDV, MyoEDV, LVESV, RVESV and MyoESV respectively. 

\subsubsection{Implementation details:} For the $\beta$-VAE-based methods, we followed the public implementation associated \footnote{\url{https://github.com/1Konny/Beta-VAE/}}. We trained for up to 1000 epochs (with a patience of 100 epochs) and used ADAM optimizer with learning rate of $5\mathrm{e}{-5}$ For the SIVAE-based methods, we followed the public implementation\footnote{\url{https://taldatech.github.io/soft-intro-vae-web/}} associated with the publication \cite{Daniel:2021}. We trained for up to 750 epochs (with a patience of 100 epochs), and we used two ADAM optimizers (one for the encoder and the decoder) with learning rate of $2\mathrm{e}{-4}$. For all methods, the size of the latent space was fixed to 128 dimensions and we trained with a batch size of 128. The hyperparameters were chosen empirically and detailled in Appendix A. An ablation study of the influence of the weight given to the attribute regularization is also conducted. More details on the implementation are available on the code repository: ***************.

\section{Experiments and results}

We compared the proposed method \gls{arsivae} to evaluate the reconstruction performance and the interpretability of the learned representation against $\beta$-VAE, SIVAE and Attri-VAE. The $\beta$-VAE and SIVAE methods do not include attribute regularization, so will be used as baselines for the reconstruction performance. Attri-VAE \cite{Pati:2021} is considered as the reference for attribute regularization and interpretability of the latent space.

\subsection{Reconstruction performance}

\begin{figure}[t]
	\centering
	\includegraphics[width = 1\textwidth]{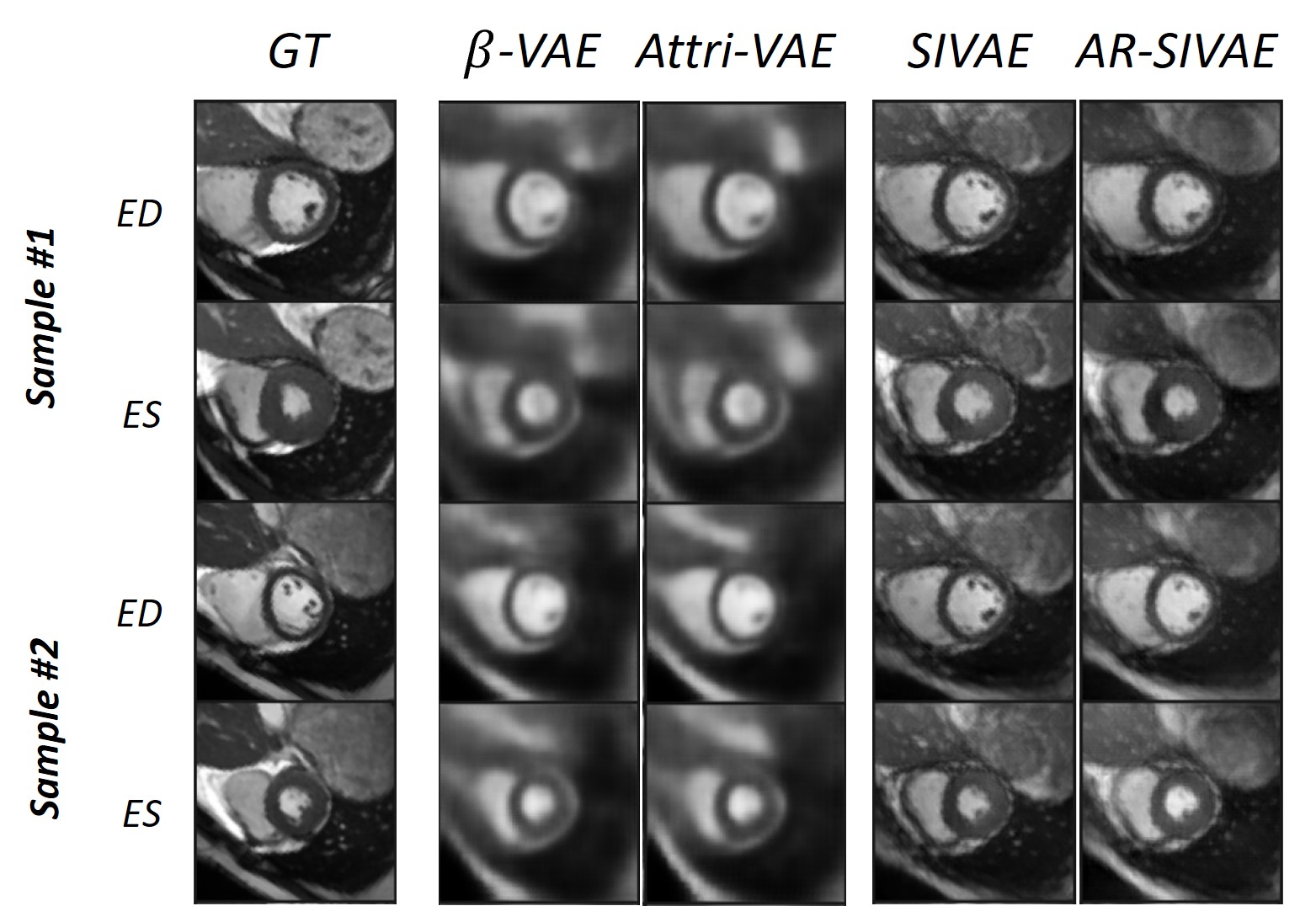}
        \caption{Qualitative evaluation of the reconstruction of two samples at \gls{es} and \gls{ed} for $\beta$-VAE, \gls{sivae}, Attri-VAE and the proposed method AR-SIVAE. The first column corresponds to the ground-truth (GT). The SIVAE-based methods ($4^{th}$ and $5^{th}$ columns) overcome the blurry reconstruction of VAE-based methods ($2^{nd}$ and $3^{rd}$ columns).}
	\label{fig:rec}
\end{figure}

We first assessed the reconstruction performance of the compared methods. We experimented using as input only the \gls{es} or \gls{ed} images, or both at the same time as two different channels. Figure \ref{fig:rec} illustrates the capacity of the SIVAE-based methods to overcome the blurry reconstruction of VAE-based methods for two samples (\gls{ed} and \gls{es} are considered together and displayed for each sample). We also report in Table. \ref{tab:rec_performance} the \gls{ssim} and \gls{lpips} metrics (the latter is a measure of the perceptual similarity between two images) when considering ED and ES together (denoted as All) and individually. Both of the VAE- and SIVAE-based methods achieved similar performance with and without attribute regularization, suggesting that the addition of the regularization term has minimal influence on reconstruction quality. Despite obtaining blurry reconstructions (illustrated in Fig. \ref{fig:rec}), the VAE-based methods achieved slightly higher \gls{ssim} scores. While it is a widely used metric to assess the similarity between two images, the \gls{ssim} often fails to detect nuances of human perception \cite{zhang2018unreasonable}. We employ the \gls{lpips} metric to address this limitation, revealing a significant improvement for the SIVAE-based methods. Furthermore,  as expected, the reconstruction of shape variability is more challenging for \gls{es} than for \gls{ed}, resulting in lower performance globally for \gls{es}. Finally, in contrast to the VAE-based methods, reconstructing ED and ES together has very little impact on the metrics compared to reconstructing individually them  when using the SIVAE-based methods.

\begin{table}[t!]
	\centering
        \caption{Evaluation in terms of reconstruction performance of the compared methods with or without attributed regularization (Reg. column) when trained using only ES, only ED or both at the same time (ED/ ES column).}
	
\begin{tabular}{p{0.15\textwidth}m{0.075\textwidth}>{\centering}m{0.15\textwidth}>{\centering}m{0.1\textwidth}>{\centering}m{0.1\textwidth}>{\centering}p{0.02\textwidth}>{\centering}m{0.15\textwidth}>{\centering}m{0.1\textwidth}>{\centering\arraybackslash}m{0.1\textwidth}}
        
		\toprule
		&  &  \multicolumn{3}{c}{\textbf{SSIM} $\uparrow$}  & & \multicolumn{3}{c}{\textbf{LPIPS} $\downarrow$}  \\
        \cmidrule{3-5}
        \cmidrule{7-9}
         & Reg. &  \multicolumn{1}{c}{All (ED/ES)} & ED & ES & & \multicolumn{1}{c}{All (ED/ES)} & ED & ES \\
		\midrule
    $\beta$-VAE & \multicolumn{1}{c}{\xmark} & \textbf{0.51}/ \textbf{0.44} & \textbf{0.54} & \textbf{0.48} & & 0.36/ 0.41 & 0.34 & 0.38 \\
    Attri-VAE  & \multicolumn{1}{c}{\cmark} & \textbf{0.51}/ \textbf{0.44} & \textbf{0.54} & \textbf{0.48} & & 0.37/ 0.42 & 0.34 & 0.38 \\
	\midrule
		SIVAE & \multicolumn{1}{c}{\xmark} & 0.46/ 0.40 & 0.45 & 0.36 & &  \textbf{0.17}/ \textbf{0.18} & 0.19  & 0.21  \\
		\textbf{AR-SIVAE}  & \multicolumn{1}{c}{\cmark} &  0.47/ 0.40 & 0.46   & 0.40 & & \textbf{0.17}/ 0.20 &\textbf{0.17}  & \textbf{0.18}  \\
		\bottomrule
\end{tabular}

	\label{tab:rec_performance}
\end{table}

\subsection{Interpretability of the latent space}

\begin{figure}[t]
	\centering
	\includegraphics[width = 1\textwidth]{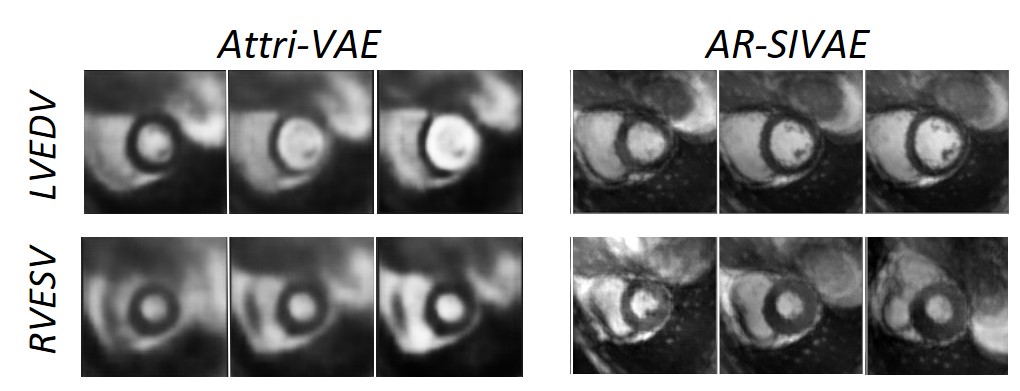}
        \caption{Walk in the regularized latent dimensions of LV end-diastolic volume (first row) and RV end-systolic volume (second row) for Attri-VAE (left) and AR-SIVAE (right).}
	\label{fig:interp}
\end{figure}

In this section, we evaluate the interpretability of the latent space when reconstructing \gls{es} and \gls{ed} together. Table. \ref{tab:rec_performance} shows the \textit{Interpretability} score, which measures the ability to predict a given attribute using only one dimension of the latent space; the \gls{sap}, which calculates the difference in \textit{Interpretability} score between the two most predictive dimensions; the \textit{Modularity} metric \cite{Ridgeway:2018}, which quantifies whether each dimension of the latent space depends on only one attribute; and the \gls{scc} which is the maximum value of the Spearman’s correlation coefficient between an attribute and each dimension of the latent space. We observed that adding the regularization to VAE- and SIVAE-based methods improved the performance of SCC (improvement of 0.15 for VAE and 0.14 for SIVAE) and \textit{Interpretability} score (improvement of 0.4 VAE and 0.25 for SIVAE). Compared to Attri-VAE, AR-SIVAE achieves better results for the SCC metrics and similar results for the \textit{Modularity} and \textit{Interpretability} scores, in particular for \gls{es} attributes. By regularizing correlated attributes such as the cardiac volumes at \gls{es} and \gls{ed}, several regularized dimensions are able to predict another attribute and this results in a lower \gls{sap} score. Figure \ref{fig:interp} illustrates a walk in the latent dimensions showing the evolution of the regularized attribute. The proposed method is able to generate non-blurry samples contrary to Attri-VAE,  while still displaying the variation of the considered attributes.

\begin{table}[t]
	\centering
 	\caption{Assessment of the interpretability of the latent space using the Spearman’s Correlation Coefficient (SCC), the \textit{Modularity} score (Mod.), and the mean \textit{Interpretability} score (Interp.). For the latter, results are shown for all attributes, as well as for the EDV- and ESV-specific ones. All the metrics are between 0 and 1, with 1 being the best performance.}
	
\begin{tabular}{p{0.15\textwidth}m{0.075\textwidth}>{\centering}m{0.1\textwidth}>{\centering}m{0.1\textwidth}>{\centering}m{0.1\textwidth}>{\centering\arraybackslash}m{0.25\textwidth}}
		\toprule
		& \multirow{2}{*}{\textbf{ Reg.}} &   \multirow{2}{*}{  \textbf{ SCC. }  } & \multirow{2}{*}{  \textbf{ Mod. }  } &  \multirow{2}{*}{  \textbf{ SAP }  }  & \multicolumn{1}{c}{\textbf{ Interp.} }  \\
            \cline{6-6}
        & & & & & All (EDV/ ESV)\\
        \midrule
	
		$\beta$-VAE & \multicolumn{1}{c}{\xmark} & 0.67  & 0.80	&  0.30 & 0.48 (0.51/ 0.48)\\
            Attri-VAE & \multicolumn{1}{c}{\cmark} &  0.82  & 0.85 	& 0.09 & \textbf{0.88} (\textbf{0.89}/ \textbf{0.87}) \\		
            
		\midrule
		  SIVAE & \multicolumn{1}{c}{\xmark} & 0.78 & \textbf{0.86}& \textbf{0.40} & 0.60 (0.62/ 0.58)  \\
		\textbf{AR-SIVAE}  & \multicolumn{1}{c}{\cmark} & \textbf{0.92} & \textbf{0.86} & 0.06 & 0.85 (0.88/ 0.82)  \\
	\bottomrule
\end{tabular}

	\label{tab:interp_performance}
\end{table}

\section{Conclusion}

This paper introduces the Attribute regularized Soft Introspective Variational Autoencoder (AR-SIVAE), which combines attribute regularization with the SIVAE framework to enhance the interpretability of the latent space while improving image generation/reconstruction capabilities. We demonstrated its effectiveness in overcoming the issue of blurry generation inherent in VAE-based methods. Nonetheless, the proposed method is limited by having a large number of hyperparameters, which makes it challenging to achieve convergence. Future efforts will focus on extending attribute regularization to non-morphometric attributes and leveraging the interpretable latent space in downstream cardiac MRI classification tasks.

\section*{Acknowledgement}

This research has been conducted using the UK Biobank Resource under Application Number 87065 and was supported by the German Federal Ministry of Health on the basis of a decision by the German Bundestag, under the frame of ERA PerMed. C.I.B. is in part supported by the Helmholtz Association under the joint research school “Munich School for Data Science - MUDS”.


\bibliographystyle{splncs04}
\bibliography{refs}

%
%
%
%
%
%
%
%
\end{document}